\documentclass[graybox]{svmult}

\usepackage[utf8]{inputenc}
\usepackage{booktabs}
\usepackage[table,xcdraw]{xcolor}
\usepackage{multirow}
\usepackage{xspace}
\usepackage{color}
\usepackage{comment}
\usepackage{graphicx}

\newcommand*{\japis}{Java I/O APIs\xspace}
\newcommand*{\japi}{Java I/O API\xspace}

\newcommand{\authorbio}[1]{
  \setlength\parindent{0pt}
  \vspace{5pt}
  #1
}

\newcommand*{\RQone}{
(RQ1) \emph{To what extent can we improve the energy efficiency of an application by statically replacing Java collections implementations?}\xspace}
\newcommand*{\RQtwo}{
(RQ2) \emph{Are the recommendations device-independent?}\xspace}
\newcommand*{\RQthree}{
(RQ3) \emph{How much does workload size impact the energy efficiency of a Java collection implementation?}\xspace}
\newcommand*{\RQfour}{
(RQ4) \emph{Are the recommendations profile-independent?}\xspace}

\newcommand*{\ct}{\textsf{CT+}\xspace}
\newcommand*{\dell}{\textbf{dell}\xspace}
\newcommand*{\asus}{\textbf{asus}\xspace}
\newcommand*{\serv}{\textbf{server}\xspace}
\newcommand*{\JSEVEN}{\textbf{J7}\xspace}
\newcommand*{\TAB}{\textbf{Tab4}\xspace}
\newcommand*{\SEIGHT}{\textbf{S8}\xspace}
\newcommand*{\GTWO}{\textbf{G2}\xspace}

\newcommand*{\pbig}{\texttt{big}\xspace}
\newcommand*{\pmedium}{\texttt{medium}\xspace}
\newcommand*{\psmall}{\texttt{small}\xspace}

\newcommand{\dacapo}{Dacapo}

\title{Small Changes, Big Impacts: Leveraging Diversity to Improve Energy Efficiency}
\author{Wellington Oliveira, Hugo Matalonga, Gustavo Pinto, \\ Fernando Castor and Jo{\~a}o Paulo Fernandes}
\institute{Federal University of Pernambuco, Federal University of Par\'{a}, University of Coimbra, Minho University}
\date{June 2020}
\begin{document}

\maketitle

\section{Introduction}

In 2012, information and communication technology was estimated to be responsible for 4.7\% of the world's electrical energy consumption~\cite{Gelenbe:2015:IIT}. Although that energy is to a large extent used to reduce energy consumption in other productive sectors~\cite{Coroama:2009:ECE,Gelenbe:2015:IIT}, it is still a considerable percentage. Moreover, that figure is estimated to grow to between 8 and 21\% of the global demand for energy by 2030~\cite{Andrae:2015:GEU}. In addition, energy has a high cost for many organizations~\cite{Andrews:2016:EUB}. Reducing that cost, even by a small percentage, can mean savings in the order of millions of dollars. 

High energy consumption also has a direct impact in our daily lives, specially when we consider mobile devices. Long battery life is considered one of the most important smart phone features by users~\cite{FRichter2018,thorwartConsumer}. In addition, from a sustainability standpoint, batteries that last longer need to be recharged less, which also increases the lifespan of mobile devices. Making the battery last longer with a single charge involves a combination of energy-efficient hardware, infrastructure software, and applications.

In the last few years, a growing body of research has proposed methods, techniques, and tools to support developers in the construction of software that consumes less energy. These solutions leverage diverse approaches such as version history mining~\cite{Hindle:2012:MSR}, analytical models~\cite{DiNucci:2017}, identifying energy-efficient color schemes~\cite{Vasquez:2018:MOO}, and optimizing the packaging of HTTP requests~\cite{Li:2016:AEO}.

In this chapter, we present a complementary approach. 
We advocate that developers should leverage software diversity to make software systems more energy-efficient. Our main insight is that non-specialists can build software that consumes less energy by alternating at development time between readily available, diversely-designed pieces of software implemented by third-parties. These pieces of software can vary in nature, granularity, and quality attributes. Examples include data structures and constructs for thread management and synchronization. 

Diversity can be leveraged in a number of different situations to improve the quality of both software systems and the processes through which they are built. According to the Merrian-Webster dictionary, diversity is ``the quality or state of having many different forms, types, ideas, etc.''. In the context of fault-tolerant software, \emph{design diversity} has been employed since the 70s~\cite{Avizienis:1984:FTD,Randell:1975:SSS}. The idea is that different implementations built from the same specification are likely to fail independently and thus can be combined to build more reliable software. Another flavor of design diversity aiming to improve reliability can be observed when developers write detailed behavioral contracts for functions~\cite{Leino:2017:ASV}. A contract can be seen as a diverse implementation written in a declarative language that is close to mathematics. 
Design diversity is also important for the construction of software systems that have dependencies on external libraries, components, or frameworks. In 2016, the un-publication of a small npm Javascript package\footnote{\url{http://www.theregister.co.uk/2016/03/23/npm\_left\_pad\_chaos/}} broke thousands of client projects. Availability of diverse packages with similar functionality can help reduce the impact of this kind of problem.
Diversity is applicable beyond software design, in other software-related situations. Not long ago, Google discussed~\cite{Dean:2003:TS} one its approaches to reduce latency: to have multiple servers serve the same request. In this scenario, we have \emph{latency (or timing) diversity}, since a multitude of factors can affect the response time of each server at any given moment. 

In this chapter we discuss the use of software diversity as a tool in the developers' toolbox to build more energy-efficient software. Diversity, in this case, expands the design and implementation options~\cite{Baldwin:2000:DRV} available for developers.
To assess the impact of these options, throughout this chapter, we revisit the main findings of research work we conducted in the past few years (e.g.,~\cite{Oliveira:2017,Oliveira:2019,Pinto:ICSME:2016,Lima:2019:HEE,matalonga2019greenhub,Rocha:2019:ESEM}).
Although these works target different programming languages, execution environments, and programming constructs, they share a common observation: \textbf{small changes can make a big difference in terms of energy consumption}. These changes can usually be implemented by very simple modifications, sometimes amounting to a single line of code. Nonetheless, the results can be significant. 

In our work we have, for example, refactored two Java systems, the \textsc{Tomcat} web server and the \textsc{Xalan} library for XML processing. Based on experimental results~\cite{Pinto:ICSME:2016}, we replaced most of the uses of the \texttt{Hashtable} class, which implements the \texttt{Map} interface, by uses of the \texttt{ConcurrentHashMap} class, which implements the same interface. In most of the cases, it was only necessary to modify the line where the \texttt{Hashtable} object was created. This simple reengineering effort promoted a reduction of up to 17.8\% in the energy consumption of \textsc{Xalan} and up to 9.32\% for \textsc{Tomcat}, when using the workloads of the DaCapo~\cite{Blackburn:2006} benchmark suite.

This chapter first introduces some of the aforementioned studies (Section~\ref{sec:design}). It then proceeds to present an automated approach to help developers to select potentially more energy-efficient options in situations where diversity is available (Section~\ref{sec:collections}). On the one hand, this approach works statically and experiments with have conducted show that it is able to improve the energy efficiency of real-world systems. On the other hand, for mobile devices, results vary widely (particularly due to the fragmentation of Android devices and their versions), which requires additional information and experimentation on their usage profiles. Based on this, we present a more recent initiative that aims to collect real-world usage information about thousands of mobile devices and make it publicly available to researchers and companies interested in energy efficiency (Section~\ref{sec:greenhub}). 


\section{Software Energy Consumption}

Although software systems do not consume energy themselves, they affect hardware utilization, leading to indirect energy consumption. Energy consumption $E$ is an accumulation of power dissipation $P$ over time $t$, that is, $E = P \times t$. Power $P$ is measured in watts, whereas energy $E$ is measured in joules. 
As an example, if one operation takes 10 seconds to complete and dissipates 5 watts, it consumes 50 joules of energy $E = 5 \times 10$. In particular, when talking about software energy consumption, one should pay attention to:

\begin{itemize}
  \item The hardware platform,
  \item The context of the computation,
  \item The time spent.
\end{itemize}

To understand the importance of a \emph{hardware platform}, consider an application that communicates through the network. Any commodity smartphone supports, at least, WiFi, 3G, and 4G. Some researchers observed that 3G can consume about 70\% more energy than WiFi, whereas 4G can consume about 30\% more energy than 3G, while performing the same task, on the same hardware platform~\cite{Kwon:ICSM:2013}. 

\emph{Context} is relevant because the way software is built and used has a critical influence on energy consumption. A program may impact energy  consumption of different parts of a device, for instance, the CPU, when performing CPU-intensive computations~\cite{Pinto:OOPSLA:2014}, the DRAM, when performing intensive accesses to data structures~\cite{Liu:2015:FASE}, the network, when sending and receiving HTTP requests~\cite{Chowdhury:SANER:2016}, or on OLED displays, when using lighter-colored backgrounds~\cite{Vasquez:2018:MOO}.

Finally, \emph{time} plays a key role in this equation. A common misconception among developers is that reducing execution time also reduces energy consumption~\cite{Manotas:2016:ICSE,Pinto:2014:MSR}, the $t$ of the energy equation. However, chances are that this reduction in execution time might increase energy consumption by imposing a heavier burden on the device, \emph{e.g.,} by using multiple CPUs~\cite{Lima:2019:HEE}. This in turn can increase the number of context switches and, as a consequence, might also increase the $P$ of the equation, impacting the overall energy consumption.

\subsection{Gauging energy consumption}

Power Measurement and Energy Estimation are high level approaches encompassing multiple techniques to gauge energy consumption at different levels of granularity. 
The first group of techniques makes use of power measurement hardware to obtain power samples. The main advantage of this technique is its ability to capture actual power use, possibly with high precision. Its main disadvantage, however, is that it is only possible to attribute the measured power to specific hardware or software elements indirectly. This usually requires software based techniques and energy estimation (see below). 
Many different power meters are currently available in the market. Different power meters have different characteristics. Among these characteristics, one of the most important is the sampling rate, that is, the number of samples obtained per second. 
The sample is often measured in watts, $P$ (power). 
Depending on the power meter used, the sampling rate can vary from 1 sample per second, to more than 10,000 samples per second.  The higher the sampling rate, the more accurate the power curve will be. 

The second area, energy estimation, assumes that developers do not have access to power measurement hardware and uses software-based techniques to predict how much energy an application will consume at run time. These predictions are based on mathematical models of how the different aspects of the hardware under examination consume, while accounting for their workloads. 
One example of this approach is the powertop\footnote{\url{https://01.org/powertop}} utility. This tool takes one sample per second and generates a log with these measurements. It analyzes the programs, device drivers, and kernel options running on a computer based on the Linux and Solaris operating systems, and estimates the power consumption resulting from their use. Powertop can also instrument laptop battery features in order to estimate power usage (in Watts) and battery life.

Running Average Power Limit (RAPL) interface~\cite{David:2010:MPE}, originally designed by Intel to enable chip-level power management, is widely supported in today's Intel architectures, including Xeon server-level CPUs and the popular i5 and i7. RAPL-enabled architectures monitor performance counters in a machine and estimate the energy consumption, storing the estimates in Machine-Specific Registers (MSRs). Such MSRs can be accessed by the OS, e.g, by means of the msr kernel module in Linux. RAPL is an appealing design, particularly because it allows energy/power consumption to be reported at a fine-grained level, e.g., monitoring CPU core, CPU uncore (caches, on-chip GPUs, and interconnects), and DRAM separately. Previous work has shown that RAPL estimates are precise when compared to measurements obtained by power measurement equipment~\cite{DiNucci:2017}. One drawback of this approach is the fact that programmers need a deep knowledge on how to use these low-level registers, which is not straightforward.

Liu and colleagues~\cite{Liu:2015:FASE} introduced jRAPL, a library for profiling Java programs running on CPUs with RAPL support. This library can be viewed as a software wrapper to access the MSRs. Since the user interface for jRAPL is simple, the programmer can focus her efforts on the high-level application design. For any block of code in the application whose energy/performance information is of interest to the user, she  just needs to enclose the code block with a pair of \texttt{statCheck} invocations. For example, the following code snippet attempts to measure the energy consumption of the \texttt{doWork()} method, whose value is the difference between the \texttt{beginning} and \texttt{end} variables:

{\small
\begin{verbatim}
double beginning = EnergyCheck.statCheck(); 
doWork();
double end = EnergyCheck.statCheck();
\end{verbatim}
}

As a shortcoming, the jRAPL library can only be used on desktop computers that leverage Intel CPUs. Thus, it provides little help for measuring energy consumption of mobile apps in tablets, smartphones, or smartwatches.  

In November, 2014, as part of Android 5.0, Google released the Android Power Profiler, which queries battery information from Android devices. It is currently available on every Android device since version 5.0 (which corresponds to over 85\% of all Android devices\footnote{https://www.statista.com/statistics/271774/share-of-android-platforms-on-mobile-devices-with-android-os/}).
The Android Power Profiler has many advantages over similar libraries. First, it requires no extra instrumentation. As the profiler is natively executed, no external applications are needed as well. Second, it also provides a straightforward interface to gather battery information and it does not require any setup. Third, the profiler also distinguishes battery usage in terms of the different components used on the device (e.g, WiFi, CPU, GPS). The Android Power Profiler, similarly to RAPL, is based on energy estimation.

To use the Android Power Profiler tools it is necessary to use the Android Debug Bridge (ADB)\footnote{https://developer.android.com/studio/command-line/adb}. ADB is a command-line tool that works like a communication interface, using the client-server model, where the device being used is the client and the development machine is the server. ADB allows one to install and debug apps, collect data about the device, execute automated tests, etc. For instance, the 
\texttt{adb shell dumpsys batterystats} command collects battery information and may save it into an output file. The exported file could be an input to other programs to manipulate and analyze data. A growing number of research works are taking advantage of the Android Power Profiler (e.g.,~\cite{DiNucci:2017:ICSE,DiNucci:2017,Gao:2017,Oliveira:2017}).

\section{Design Decisions}\label{sec:design}

In this section we explore three different approaches that share the same observation that small design decisions can greatly impact energy consumption. More specifically, we discuss how it is possible to reduce the energy footprint of software systems by leveraging diversity in IO primitives (Section~\ref{sec:io}), collection implementations (Section~\ref{sec:backcollections}), and concurrent programming constructs (Section~\ref{sec:concurrency}).

\subsection{IO Constructs}\label{sec:io}
 
I/O programming constructs are not only the building blocks of several low-level communication channels such as sockets or database drivers, but also the bedrock of high-level software applications that have anything to do with data storage or transmission. Despite their widespread use, the energy consumption of I/O programming constructs is not well understood. This is particularly unfortunate since related work suggests that I/O APIs could severely impact energy consumption. For instance, Lyu and colleagues~\cite{Lyu:2017} indicated that about 10\% of the energy consumption of mobile applications is spent in I/O operations. Similarly, Liu and colleagues~\cite{Liu:2015:FASE} pointed out that it was possible to save 4.29\% of energy consumption by changing I/O programming constructs. A comprehensive energy characterization of I/O programming constructs could help practitioners to further improve the energy behaviors of their software applications.

In the study of Rocha and colleagues~\cite{Rocha:2019:ESEM}, we presented a comprehensive characterization of \japis. In this work we conducted a broad experimental exploration of 22 \japis, aiming to answer two research questions: (RQ1) \emph{What is the energy consumption behavior of the \japis}? and (RQ2) \emph{can we improve the energy consumption of non-trivial benchmarks by refactoring their use of \japis}?

To answer these research questions, we employ what we consider to be three types of benchmarks. For the first research question, we created and instrumented 22 \textit{micro-benchmarks}. The \textit{micro-benchmarks} are small programs (around 200 lines of code) that perform a single task (e.g., reading a file from the disk), each one using a different \japi. These \japis have been introduced in the Java programming language in its very early versions and are in widespread use. For instance, the  \texttt{FileInputStream} \japi is used in 2,823 open source projects in BOA~\cite{Dyer:2015:BUL} (we ran this query in April 2019). Each one of the studied \japis implements at least one method for input operations or at least one method for output operations.

For the second research question, we performed refactorings in the code base of \textit{optimized benchmarks} and \textit{macro-benchmarks}. On the one hand, \textit{optimized benchmarks} are similar to micro-benchmarks in size, but are optimized for performance, while \textit{macro benchmarks} are full-fledged working software systems comprising thousands of lines of code.  The \textit{macro-benchmarks} used are the following: \textsc{Xalan} (an XSLT processor that translates XML documents into HTML files, or other types of documents), \textsc{Fop} (a XSLT processor that translates XML documents into HTML files, or other types of documents), \textsc{Batik} (a  toolkit for applications that want to use images in the Scalable Vector Graphics (SVG) format), \textsc{Commons-IO} (a utility library used to provide high-level I/O abstractions to third party software application), and \textsc{pgjdbc} (the official PostgreSQL driver for the Java programming language).
 
To avoid non-working solutions, we focused on refactorings that do not require extensive code changes (e.g., changes between \japis that extend the same interface). These refactorings could also be easily automated by a general-purpose tool. To conduct the experimentation process, we executed each benchmark 10 times. Since it requires some time for the Just-In-Time (JIT) compiler to identify the hot code and perform optimizations, we discarded the first three executions of the benchmarks. We report the average of the seven remaining executions. We also fixed the garbage collector and the heap size accordingly: we used the parallel garbage collector (\texttt{-XX:+UseParallelGC}), and the heap size was fixed at 261 MB, minimum (\texttt{-Xms}), and 4,183 MB, maximum (\texttt{-Xmx}). No other JVM options were employed.


\begin{figure*}[t]
\begin{center}
$
\begin{array}{cc}
\texttt{(a) Input} & \texttt{(b) Output} \\
\includegraphics[width=.45\textwidth, trim= 0px 10px 180px 0px]{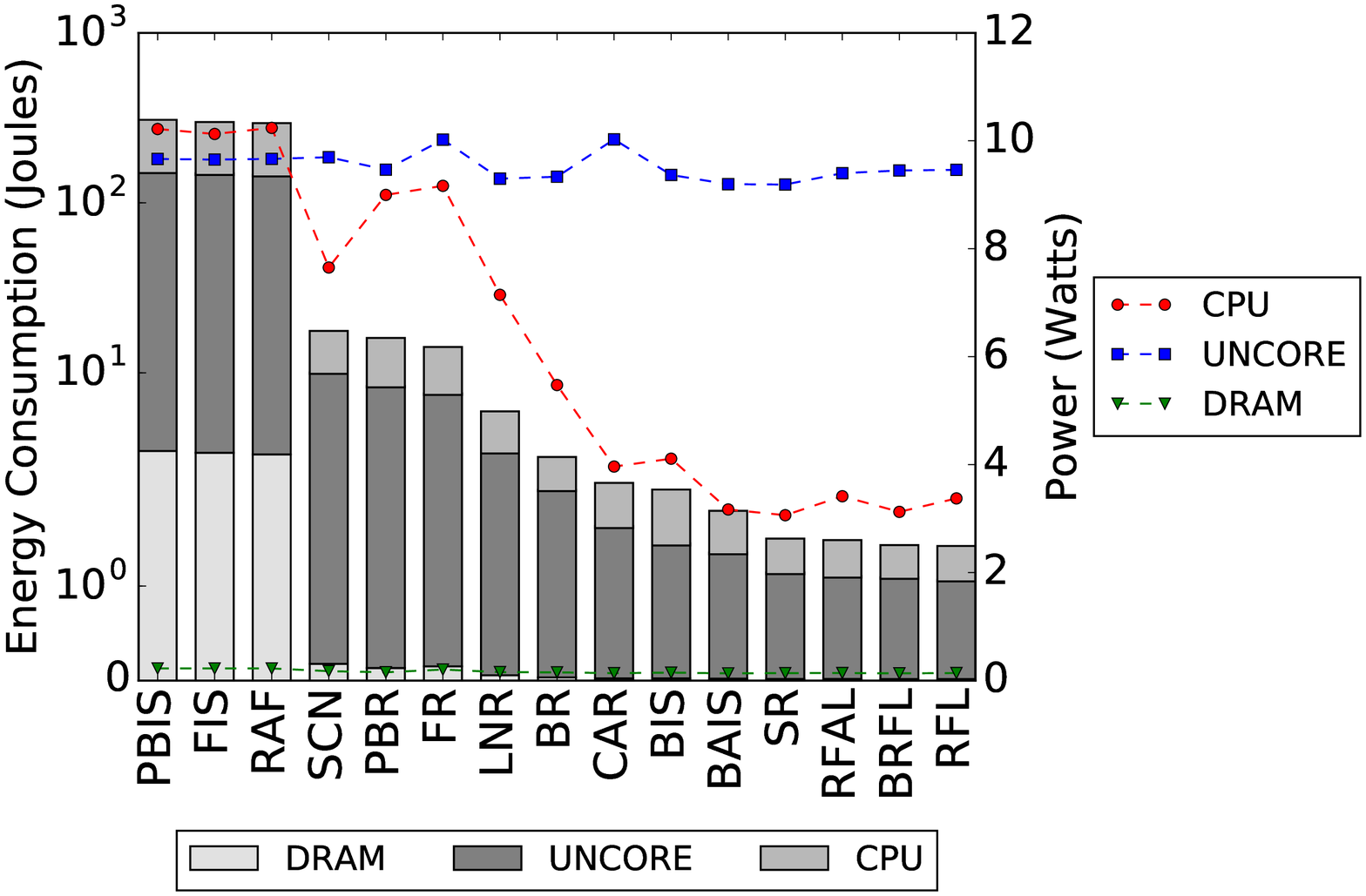} &
\includegraphics[width=.59\textwidth, trim= 0px 10px 10px 0px]{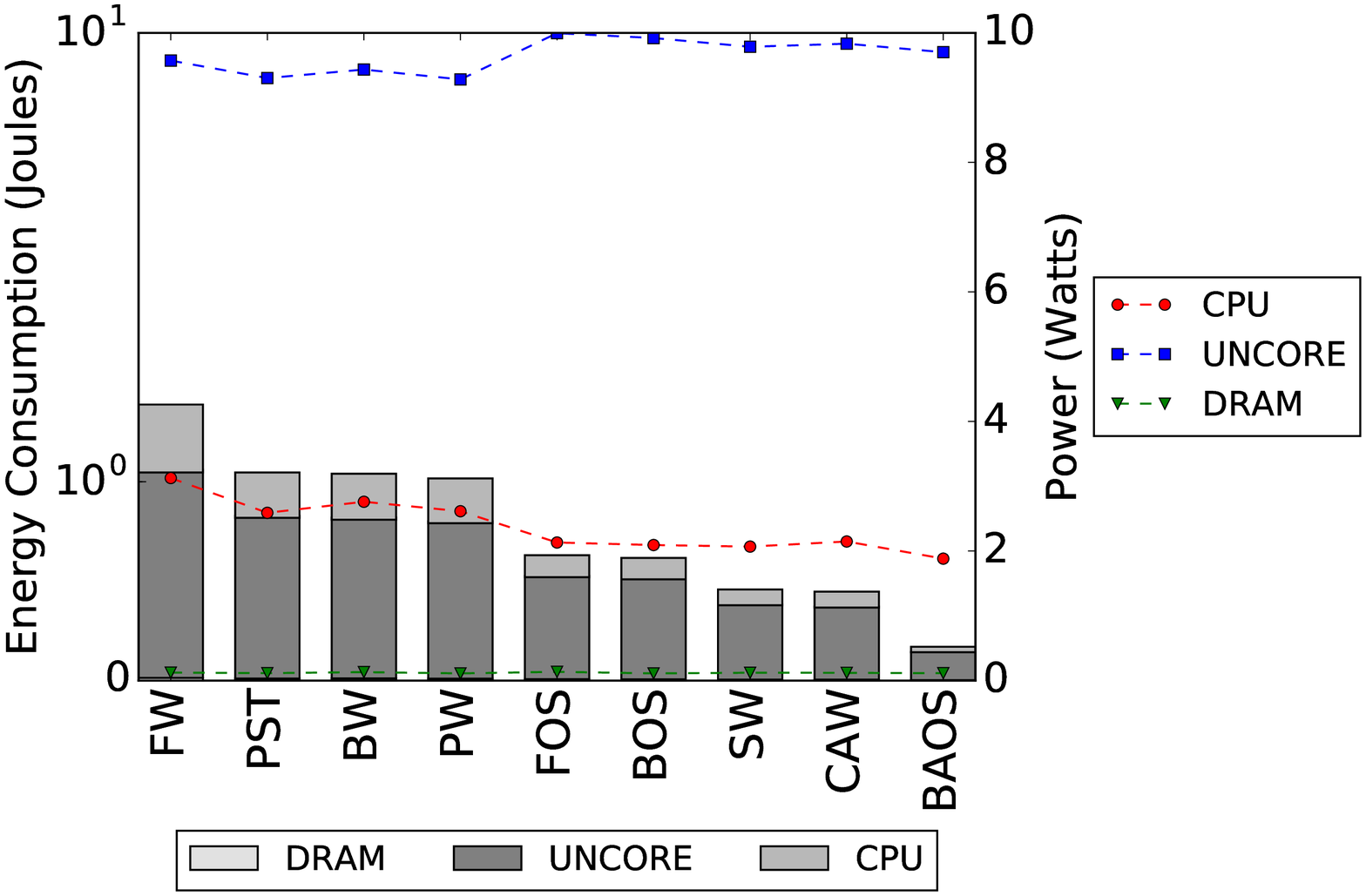} \\
\end{array}
$
\end{center}
\caption{Energy consumption behavior of \japis. Energy data is presented in a logarithmic scale. For the figure on the left, PBIS stands for \texttt{PushbackInputStream}, FIS stands for \texttt{FileInputStream}, RAF stands for \texttt{RadomAccessFile}, SCN stands for \texttt{Scanner}, PBR stands for \texttt{PushbackReader}, FR stands for \texttt{FileReader}, LNR stands for \texttt{LineNumberReader}, BR stands for \texttt{BufferedReader}, CAR stands for \texttt{CharArrayReader}, BIS stands for \texttt{BufferedInputStream}, BAIS stands for \texttt{ByteArrayInputStream}, SR stands for \texttt{StringReader}, RFAL stands for \texttt{Files.readAllLines}, BRFL stands for \texttt{Files.newBufferedReader}, and RFL stands for \texttt{Files.lines}. For the figure on the right, FW stands for \texttt{FileWriter}, PST stands for \texttt{PrintStream}, BW stands for \texttt{BufferedWriter}, PW stands for \texttt{PrintWriter}, FOS stands for \texttt{FileOutputStream}, BOS stands for \texttt{BufferedOutputStream}, SW stands for \texttt{StringWriter}, CAW stands for \texttt{CharArrayWriter}, BAOS stands for \texttt{ByteArrayOutputStream}.}
\label{fig:microbench}
\end{figure*}

After conducting this process, we observed many interesting findings. Figure~\ref{fig:microbench} shows an overview of the energy behavior of \japis for the micro-benchmarks.
First, we found that input operations consume more energy than output operations (on average: 96 joules vs 0.80 joules, respectively). The \texttt{PushbackInputStream} \japi is the most energy consuming one (492 joules consumed), followed by \texttt{FileInputStream} (474 joules). Analyzing the \texttt{PushbackInputStream} implementation, we perceived that this \japi adds a flag in the \texttt{InputStream} that marks bytes as ``not read''. Such bytes are included back in the buffer to be read again. However, before reading the bytes, this \japi also checks whether the stream is still open using the \texttt{ensureOpen()} method. This repetitive operation could be the source of this high energy consumption. The \texttt{Files} \japi, however, which could act as a potential replacement for \texttt{FileInputStream}, is the one with the least energy consumption, when executing its \texttt{lines} method (1,86 joules)

When considering the macro- and optimized-benchmarks, it was not possible to use all the \japis mentioned in Figure~\ref{fig:microbench}. This happened because there is a semantic gap between the \japis that do not inherit from the same parent, and we opted not to bridge this gap. We then only refactored instances of \japis that share the same parent class. This problem did not occur for the micro-benchmarks because of the more straightforward way in which they use the APIs. In the end, we had 21 refactored versions of these benchmarks.

Overall, when refactoring the macro- and optimized-benchmarks, we observed energy improvements in eight out of the 21 refactored versions of the benchmarks.
In particular, we observed that one optimized benchmark and one macro benchmark improved their overall energy consumption when changing their use of \japis to the \texttt{Files} class. 
With very minor modifications, we were able to improve up to 17\% of the energy consumption of these benchmarks.  These initial results provide evidence that small changes in \japis might have the potential of improving the energy consumption of benchmarks already optimized for performance.

\subsection{Collections Constructs}\label{sec:backcollections}

Collections provide easy access to reliable implementations that can reduce the complexity of developing applications.  In Java, each collection's API has multiple implementations.
Collection implementations that can be safely used by several concurrent threads are considered ``thread-safe''.
This safety usually comes with extra complexity or inferior performance, which might favor the use of ``thread-unsafe'' collections. This is expected, since there is a number of different algorithms and data structures that can implement the abstract concept of lists, sets, and maps. 
There are a number of different ways a collection can be implemented and these diverse implementations can have non-negligible impact on energy consumption.


In the last few years, a number of researchers have attempted to address the problem of helping developers to understand collections energy usage~\cite{Hasan:2016,Lima:2019:HEE,Manotas:2014:ICSE,Oliveira:2019,Pereira:2016:GREENS,Pinto:ICSME:2016}. 
These works conducted extensive exploration of collection usage. While some papers focused on the energy usage of collection implementations that are part of the Java Development Kit~\cite{Pinto:ICSME:2016}, others were broader in scope and covered not only the official implementations but also third-party libraries~\cite{Oliveira:2019}. Similarly, while some works performed the experiments on commodity devices~\cite{Pereira:2016:GREENS}, others conducted experiments on servers~\cite{Pinto:ICSME:2016}, while others also experimented with mobile devices~\cite{Hasan:2016,Oliveira:2019}. Generally speaking, these works followed a similar approach for collecting data: they created small and large benchmarks, and executed these benchmarks 10 or more times, reporting the averages as the results. 

The work of Oliveira and colleagues~\cite{Oliveira:2019} followed a sightly different approach because they created the so called ``energy profiles'', inspired by previous work~\cite{Hasan:2016}, and attempted to make recommendations by leveraging these profiles. We devote Section~\ref{sec:collections} to providing a more comprehensive overview of this work. An energy profile is a number that can be used to compare similar constructs under the same circumstances. Energy profiles for collections can be produced by executing several micro-benchmarks on different collection operations, aiming to gather information about the energy behavior of these programming constructs in an application-independent way. 
For instance, an energy profile for the operation \texttt{ArrayList.add(Object o)} could be 10.
After we create the profiles, we perform static analysis to estimate in which ways and how intensively a system employs these collections. If we know that the program under investigation uses exclusively \texttt{ArrayList.add(Object o)} 100 times, its energy consumption could be (roughly) inferred as 100 $\times$ 10 (their energy profile). 
Since the work of Oliveira and colleagues~\cite{Oliveira:2019} focuses on code recommendation, a collection is more likely to be recommended if its energy profiles are low.


Since these works performed computations in very different environments, the results cannot be easily merged together. However, some interesting findings seem to emerge. 
For instance, these papers explored the energy consumption of the most commonly-used methods. In the case of  \texttt{ArrayList}, they investigated the \texttt{add(Object o)} method. For example, in the work of Pinto and colleagues~\cite{Pinto:ICSME:2016}, the authors observed that method \texttt{ArrayList.add(Object o)} consumes the least energy, when compared to the thread-safe implementations. On the other hand, both Pinto et al.~\cite{Pinto:ICSME:2016} and Pereira et al~\cite{Pereira:2016:GREENS} observed that the most energy-consuming implementation among the thread-safe collections is  \texttt{CopyOnWriteArrayList}. In particular, Pinto et al.~\cite{Pinto:ICSME:2016} noted that insertion operations over \texttt{CopyOnWriteArrayList} consumed about 152x more energy than \texttt{Vector} (which consumes 14x more than \texttt{ArrayList}). In terms of \texttt{Map} implementations, it was found that the concurrent implementation \texttt{ConcurrentHashMap} had similar performance when compared to the non-thread safe implementation, \texttt{LinkedHashMap}, on both insertion and removal operations. Indeed, \texttt{ConcurrentHashMap} performed around three times better than \texttt{Hashtable}, one of the most common \texttt{Map} implementations.

It is important to note that these findings were observed in small benchmarks, that is, $\sim$100 lines of code programs that perform one collection operation a number of times. 
Given these observations in a controlled setting, Pinto et al.~\cite{Pinto:ICSME:2016} also manually refactored two large scale open source programs: \textsc{Xalan} (a program that transforms XML documents into HTML, which had 170k lines of code in the version we studied) and \textsc{Tomcat} (an open-source Web server, which had 188k lines of code in the version we studied). In both programs, the authors changed 100+ uses of \texttt{Hashtable} to \texttt{ConcurrentHashMap}. After applying these modifications, it was observed an energy saving of 12\% for \textsc{Xalan} and 17\% \textsc{Tomcat}, considering the workloads of the DaCapo benchmark suite~\cite{Blackburn:2006}.
Oliveira and colleagues~\cite{Oliveira:2019} also employed a similar approach for alternating between collection implementations. In their work, they found that by refactoring from \texttt{ArrayList} to \texttt{FastList} (a third-party \texttt{List} implementation) it was possible to save 17\% in energy consumption of one mobile app, \textsc{PasswordGen}. These findings share a common trend: with no prior knowledge of the application domains or the system implementations, it was possible to reduce the energy consumption of a software system by means of simple changes in collection usage.

\subsection{Concurrent Programming Constructs}\label{sec:concurrency}

Concurrency control and thread management are additional software features where it is possible to reap the benefits of software diversity. Early work by Trefethen and Thiyagalingam~\cite{Trefethen:2013:EAS} observed that, for parallel applications in the area of scientific computing, performance is often not a proxy for energy consumption. A subsequent study~\cite{Pinto:OOPSLA:2014} investigated the impact of different approaches to manage concurrent and parallel execution in Java programs. This study found out that different thread management approaches, e.g., per-core threads, thread pools, and work-stealing, have diverse, significant, and hard to predict impacts on energy consumption. It also observed that performance is not a good proxy for energy efficiency in the studied benchmarks, which comprised both small programs and real-world, high-performance Java applications. 

These studies inspired us to investigate how thread management constructs affect energy consumption in a different setting, namely, programs written in Haskell, a lazy, purely functional programming language. Haskell programs can create lightweight threads that may be associated with a specific physical core or operating system thread, or managed entirely by the Haskell scheduler. Furthermore, the language has multiple primitives for data sharing between threads which act as concurrency control primitives, including a lock-based approach, a fully-featured implementation of software transactional memory~\cite{Shavit:1997:STM} (STM), and an STM-based solution that simulates locks. 

We conducted a study with nine Haskell benchmarks. The benchmarks were selected from multiple sources, such as the Computer Language Benchmarks Game\footnote{https://benchmarksgame-team.pages.debian.net/benchmarksgame/index.html} and Rosetta Code\footnote{http://www.rosettacode.org/}. We selected the benchmarks based on their diversity. For instance, two of them are synchronization-intensive programs, two are CPU-intensive and scale up well on a multicore machine, two are CPU- and memory-intensive, one is IO-intensive, one is CPU- and IO- intensive, and one is peculiar in that it is CPU-, memory-, synchronization-, and IO-intensive. We implemented and ran different variants of these benchmarks considering the nine possible combinations of thread management constructs and data sharing/concurrency control primitives. Not every possible variant could be used, e.g., because some benchmarks do not leverage concurrency control. The details of the methodology of this study are presented elsewhere~\cite{Lima:2019:HEE}.

We ran all the experiments on a server machine with 2x10-core Intel Xeon E5-2660 v2 processors (Ivy Bridge microarchitecture, 2-node NUMA) and 256GB of DDR3 1600MHz memory. This machine runs the Ubuntu Server 14.04.3 LTS (kernel 3.19.0-25) OS. The compiler was GHC 7.10.2. The benchmarks were exercised by Criterion~\cite{criterionsite}, a benchmarking library to measure the performance of Haskell code. To collect information about energy usage, we had to modify the implementations of Criterion and the Haskell profiler to make them energy-aware. We executed the benchmarks with 1, 2, 4, 8, 16, 20, 32, 40, and 64 capabilities. The number of capabilities of the Haskell runtime determines how many Haskell threads can run truly simultaneously at any given time. 

\begin{figure}[tb]
	\centering
	\includegraphics[scale = 0.20]{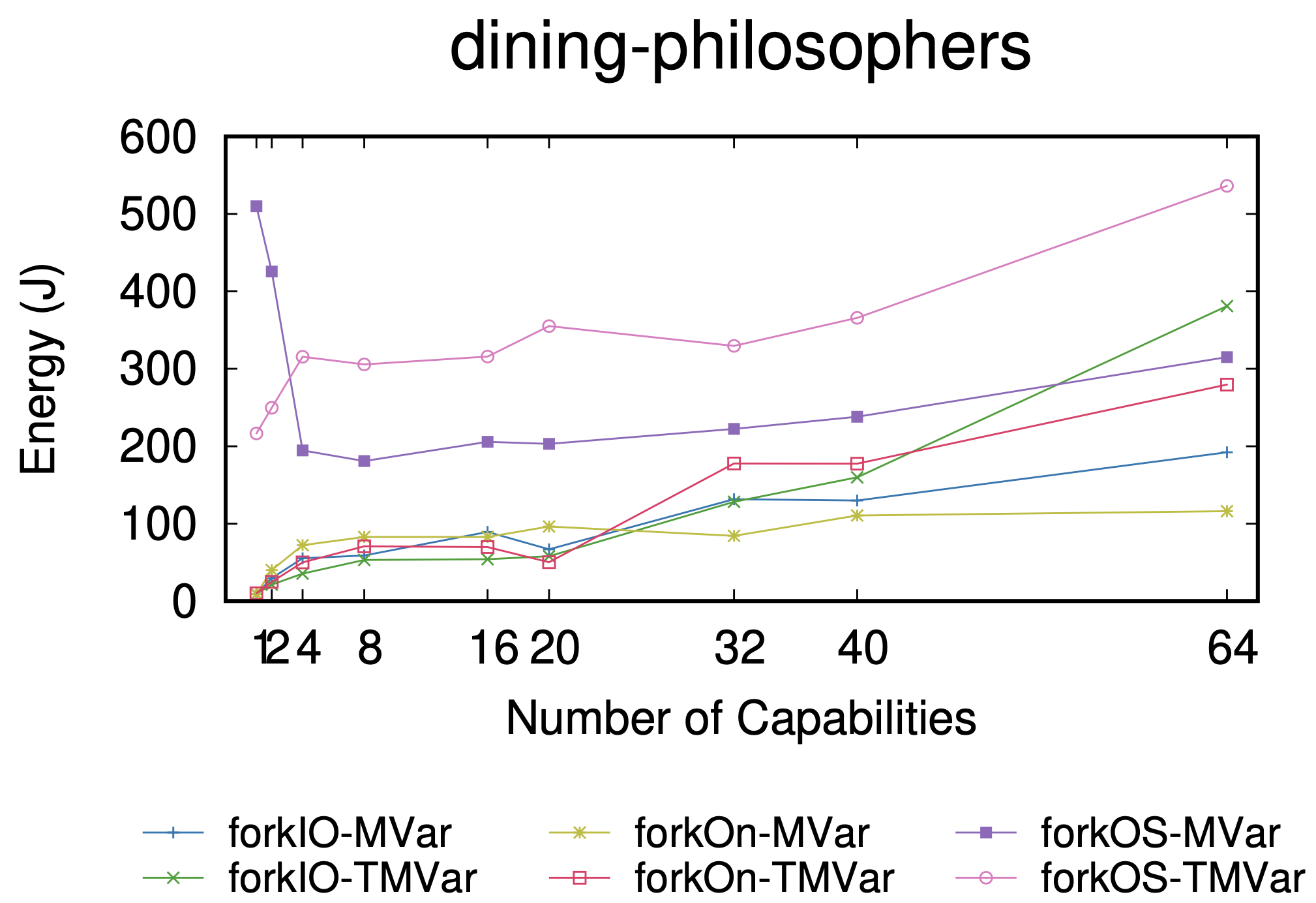}
	\caption{Energy measurements for the dining philosophers benchmark, considering six combinations of thread management constructs (\texttt{forkIO}, \texttt{forkOn}, and \texttt{forkOS}) and concurrency control primitives (\texttt{MVar} and \texttt{TMVar})~\cite{Lima:2019:HEE}}
	\label{img:dining}
\end{figure}

Once again, we found that small changes can make a big difference in terms of energy consumption. For example, in one of our benchmarks, under a specific configuration, choosing one data sharing primitive (\texttt{MVar}) over another (\texttt{TMVar}) can yield 60\% energy savings. Notwithstanding, there is no universal winner. The results vary depending on the characteristics of each program. In another benchmark, \texttt{TMVar}s can yield up to 30\% energy savings over \texttt{MVar}s. Figure~\ref{img:dining} illustrates an extreme case. When considering 20 capabilities, the \texttt{forkOS-TMVar} variant of the dining philosophers benchmark consumed 268\% more energy than the \texttt{forkIO-MVar} variant. These results indicate that it is also possible exploit software diversity in Haskell in order to improve energy efficiency.

\begin{figure}[tb]
	\centering
	\includegraphics[scale = 0.20]{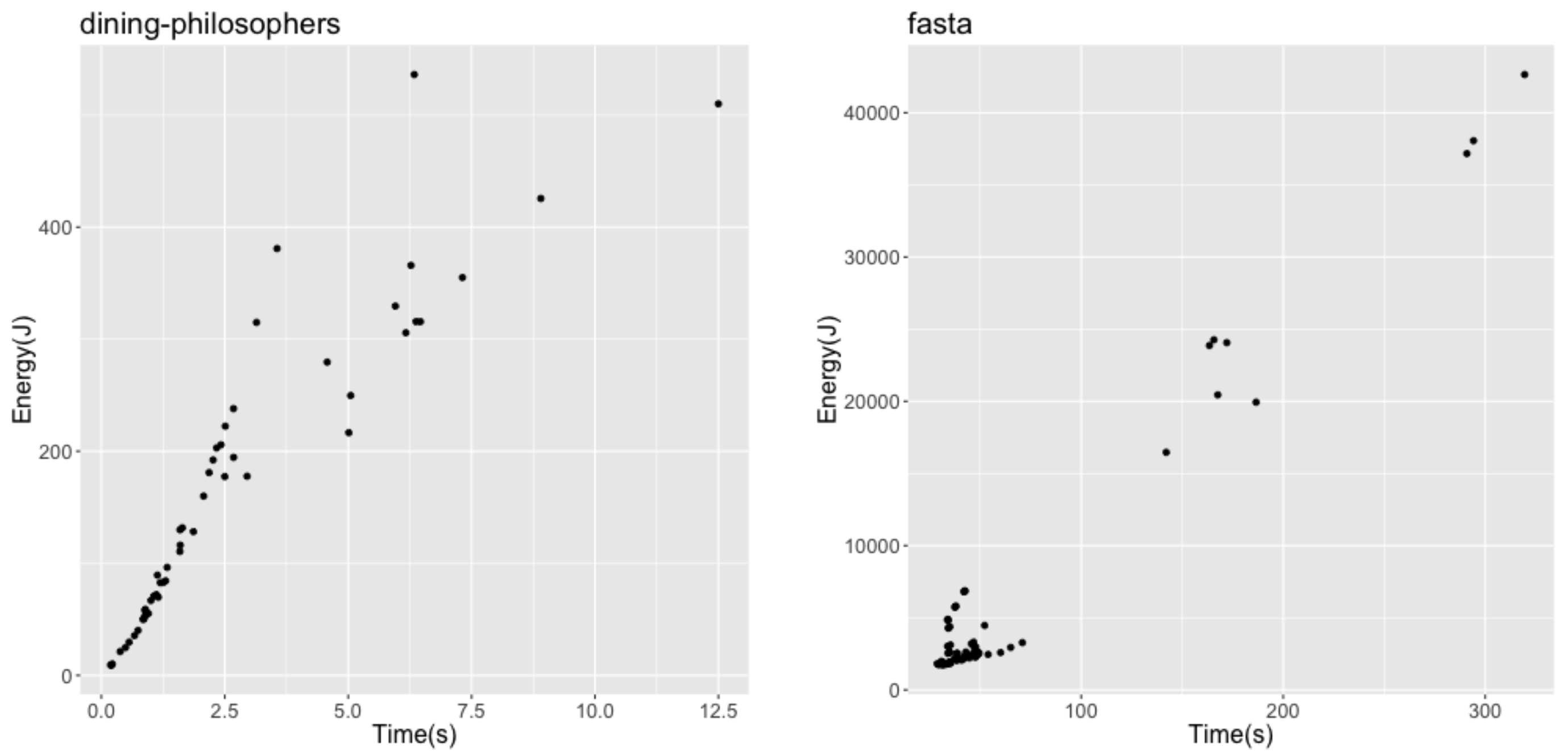}
	\caption{Scatterplots for the relationship between time (x-axis) and energy (y-axis) for two of the analyzed Haskell benchmarks~\cite{Lima:2019:HEE}.}
	\label{img:scatterplots}
\end{figure}

Similarly to previous studies~\cite{Pinto:OOPSLA:2014,Trefethen:2013:EAS}, we found out that the relationship between energy consumption and performance is not always clear.
High performance is usually a proxy for low energy consumption. Nonetheless we found scenarios where the configuration with the best performance (30\% faster than the one with the worst performance) also exhibited the second worst energy consumption (used 133\% more energy than the one with the lowest usage). The scatterplots in Figure~\ref{img:scatterplots} illustrate how energy and time are imperfectly aligned. This is different from what we would observe, for example, in sequential Haskell collections~\cite{Lima:2019:HEE}, where the points would be almost perfectly aligned along the diagonal.

In addition to these results, we propose some guidelines that Haskell developers can follow to make applications more energy efficient, based on our empirical results. First, CPU-bound applications should avoid setting more capabilities than physical cores, since these applications will not benefit from the enhanced thread switching afforded by hyperthreading\footnote{https://www.intel.com/content/www/us/en/architecture-and-technology/hyper-threading/hyper-threading-technology.html} and similar approaches. Second, they should use the \texttt{forkOn} function, which attempts to pin threads to specific physical cores, to create threads in embarassingly parallel applications. This reduces thread migration overhead in applications where workloads are evenly distributed among threads and threads do not have data dependencies. Third, they should avoid using the \texttt{forkOS} function to spawn new threads. Since this function binds a Haskell thread to an OS thread, it results in thread switching involving OS threads whenever new Haskell threads must be executed. Fourth, if energy matters, only use STM if transaction conflicts are rare. Although transactional memory may improve performance due to optimistic concurrency, the large number of conflicts can have a strong impact on energy consumption, even if they do not hinder performance. As mentioned before, for one of the analyzed benchmarks, the variant with the best performance exhibited more than twice the energy consumption of the most energy-efficient variant. 

\section{Recommending Java Collections}\label{sec:collections}


Developing applications, complex systems that have functional and non-functional requirements combined to solve a non-trivial problem, can be a difficult task. 
Leveraging previously implemented software solutions to solve pieces of these challenging problems can help to reduce that complexity. Examples of this solutions are libraries, APIs, frameworks, gists\footnote{\url{http://gist.github.com}}, and answers from Q\&A sites such as StackOverflow.

In this context, designing and implementing software has become a task of selecting appropriate solutions among multiple options~\cite{Baldwin:2000:DRV} and combining them to build working systems. 
We call \emph{energy variation hotspots} the programming constructs, idioms, libraries, components, and tools in a system for which there are multiple, interchangeable, readily-available solutions that have potentially different energy footprints. 
A number of previous papers have measured and analyzed different types of energy variation hotspots, such as  programming languages \cite{Georgiou:2020,Oliveira:2017,Pereira:2017}, API usage \cite{Aggarwal:2014:PSC,Vasquez:2018:MOO,Rocha:2019:ESEM}, thread management constructs \cite{Lima:2019:HEE,Pinto:OOPSLA:2014}, data structures \cite{Hasan:2016,Lima:2019:HEE,Pereira:2016:GREENS,Pinto:ICSME:2016}, color schemes~\cite{Li:2014:MWA,Linares-Vasquez:2015},
and machine learning approaches~\cite{Hindle:2019}, among many others. 
Having to choose the most energy efficient solution for an energy variation hotspot can be difficult for developers, as the energy consumption of these constructs usually is not easily measurable. 
Furthermore, information on how to execute tests to measure the energy impact of different solutions can be hard to find.        


In this section, we present our solution to reduce the energy consumption of software applications, making it easier for non-specialist developers to exploit energy variation hotspots.
This solution can be separated in three different steps:
On the first step, we exercise the the available alternative solutions, aiming to build energy consumption profiles~\cite{Hasan:2016}.
On the second step, we analyze the application, collecting and organizing the usage of the selected energy variation hotspots, in particular, to estimate how intensively the system uses them. 
Finally, on the third step, we combine the energy profile and the results of analyzing the system to make potentially energy-saving recommendations specific to the application-device pair. 
This approach is instantiated in an energy-saving tool named \ct.
Using this tool, non-specialist developers can optimize the energy-efficiency of Java collections.

While experimenting with \ct, we selected collections from three different sources: Java Collections Framework (JCF)\footnote{\url{https://docs.oracle.com/javase/8/docs/technotes/guides/collections/}}, Apache Commons Collections\footnote{\url{{https://commons.apache.org/proper/commons-collections/}}}, and Eclipse Collections\footnote{\url{{https://www.eclipse.org/collections/}}}.
These sources are widely used on Java projects, with a query on GitHub projects\footnote{These queries were executed in April 2020} showing 1,276,939 occurrences for Apache Commons, 537,956 occurrences for Eclipse Collections and 85,865,270 occurrences for the most widely used collection implementations from the JCF. 
All sources have thread-safe and thread-unsafe collections.

We implemented \ct following our approach step-by-step:
In the first step, it automatically runs multiple micro-benchmarks (i.e., executing specific collection operations such as \texttt{List.add(Object o)}) for 39 distinct Java collection implementations in an application-independent manner and builds their energy profiles. 
\texttt{List}, \texttt{Map}, and \texttt{Set} are the three collection APIs targeted by our tool.
A varying number of operations was exercised for each API, with \texttt{List} having 12 operations, \texttt{Map} 4, and \texttt{Set} 3.
Our collection pool comprises implementations from the Java Collections Framework (25 implementations), Apache Commons Collections (5 implementations), and Eclipse Collections (9 implementations).
The energy consumption profile is built with the data from these micro-benchmarks.

For the second step, an inter-procedural static analysis using WALA\footnote{\url{{http://wala.sourceforge.net/wiki/index.php/Main\_Page}}} is performed on the application source code.
This analysis collects and organizes data about how the application uses collection implementations on its souce code, such as frequency and location of use, which operations were used, method and variable names, calling context, among others.

The third and final step consists of combining these two pieces of information, that is, the energy profile and the analysis of the application. \ct identifies the most energy-efficient collection implementations across the whole program and automatically applies these recommendations to the source code. 
We evaluated \ct in two distinct studies, analyzing the impact of different devices and of different energy profiles, aiming to answer the following four research questions:\RQone,\RQtwo,\RQthree, and\RQfour


\subsection{Evaluation}

The main objective of our evaluation was to compare the energy consumption of the original versions of software systems with the versions where the recommendations made by \ct were applied. This was made across two different studies.
Overall, our evaluation comprises two different execution environments, \textbf{desktop} and \textbf{mobile}, and six distinct devices.

In the first part of our experiments, our main goal was to evaluate the collection implementation recommendations made by \ct (RQ1 and RQ2). 
On the desktop environment we executed \ct across two machines, a notebook and a high-end server. We labeled the notebook \dell (Dell Inspiron 7000) and the server as \serv (the same machine described in Section~\ref{sec:concurrency}). 
On the mobile environment, we executed our tool on three smartphones and a tablet: Samsung Galaxy J7 (\JSEVEN), Samsung Galaxy S8 (\SEIGHT), Motorola G2 (\GTWO), and Samsung Galaxy Tab 4 (\TAB). 
In this experiment, we analyzed seven desktop-based software systems: \textsc{Barbecue}, \textsc{Battlecry}, \textsc{JodaTime}, version 6.0.20 of \textsc{Tomcat}, \textsc{Twfbplayer}, \textsc{Xalan}, and \textsc{Xisemele}; two mobile-based software systems: \textsc{FastSearch} and \textsc{PasswordGen}; and three that work on both environments: \textsc{Apache Commons Math 3.4} (\textsc{Commons Math} for short), \textsc{Google Gson}, and \textsc{XStream}.

In the second part of the experiment, we have analyzed the energy impact of three different strategies to build energy profiles (RQ3 and RQ4)
For this study, a single device was used: a notebook \asus (ASUS X555UB). 
To explore the impact of different energy profiles on the energy-efficiency behavior of Java collection implementations, we created three different profiles for~\asus: \psmall, \pmedium, and \pbig. 
These profiles were created to simulate three different scenarios of usage intensity of the collection implementations:
\psmall, to be used for applications that have a light usage of collections;
\pmedium, an intermediate profile, for general purpose usage;
and 
\pbig to be used on applications that have a very intense usage of collections.
We used as targets systems six applications from the latest version of \dacapo: \textsc{Biojava}, \textsc{Cassandra}, \textsc{Graphchi}, \textsc{Kafka}, \textsc{Zxing}, and version 9.0.2 of \textsc{Tomcat}. The latter with two different types of workload: \textsc{large} and \textsc{huge}.

To measure the energy consumption of the devices on both studies, we employ jRAPL to collect the energy data in the desktop environment and the Android Energy Profiler in the mobile environment.
While running our experiments, for some systems the difference between original and modified versions was not statistically significant. 
That was the case for \textsc{Twfbplayer}, \textsc{Xisemele}, \textsc{Cassandra}, and \textsc{Kafka} and for the device \TAB.
To better focus on more relevant data, these results are not be presented here. 
However, all data from every system used in our experiments can be found in the companion website, at \textbf{\url{https://energycollections.github.io/}}.

\begin{figure}[tb]
	\centering
	\includegraphics[scale = 0.75, clip = true, trim= 1px 0px 0px 0px]{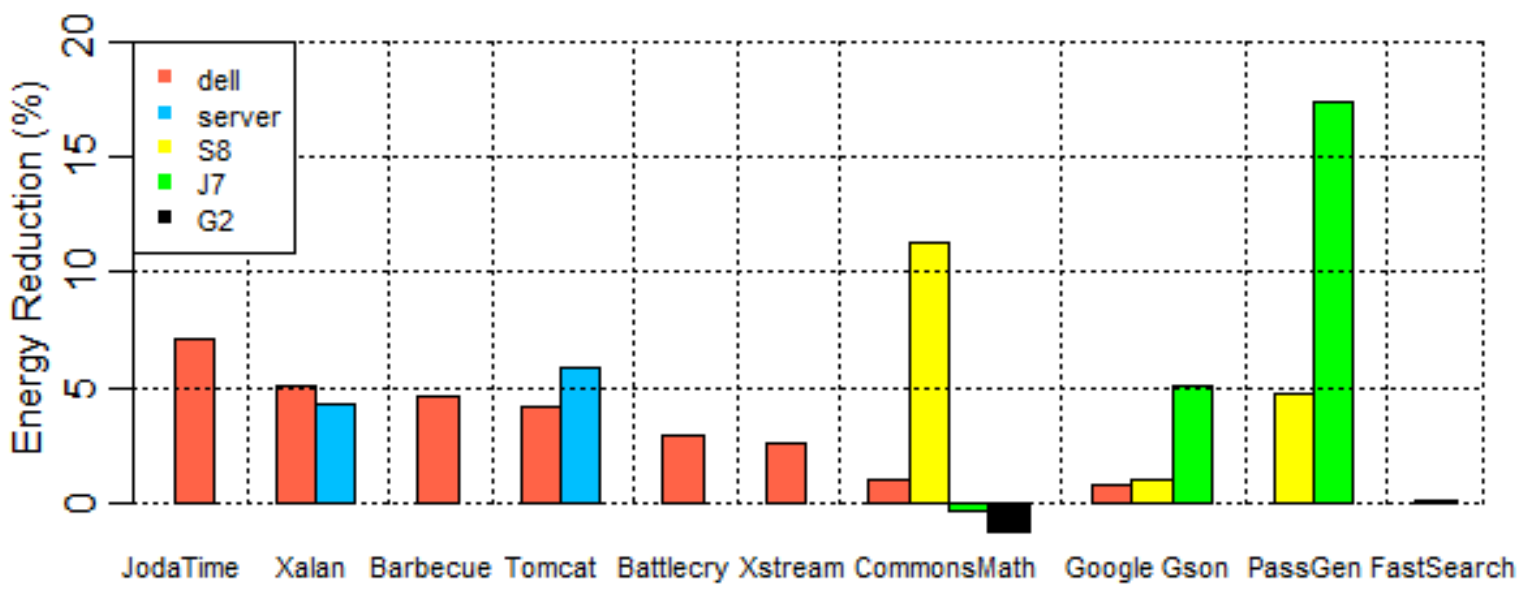}
	\caption{Percentage of energy reduction on the study about different devices. Greater is better.}
	\label{img:resultsDevices}
\end{figure}

Across both studies, \ct performed 1,454 changes that impacted the energy consumption across 17 software systems, 12 targeting a desktop environment, 2 targeting a mobile environment, and 3 that work in both scenarios, for a total of 46 modified versions. The analyzed applications were, for the most part, mature systems comprising thousands of lines of code (LoC), such as  \textsc{BioJava} with 914kLoC,
\textsc{Cassandra} with 466kLoC, and
\textsc{Tomcat} with 433kLoC. 
Even without any prior knowledge of the application domains, \ct reduced the energy consumption of 13 out of the 17 systems. 
\\[0.6cm]
\noindent
\textbf{Analyzing different devices}.
Figure~\ref{img:resultsDevices} summarizes the results of our modifications on the desktop and mobile devices.
For the desktop environment, \ct made 477 recommendations, with all the modified systems consuming less energy than the original versions.  
Among the software systems that only ran on the \dell machine, \textsc{JodaTime} exhibited the greatest improvement, with the modified version consuming 7.13\% less than the original one. 
The modified version of \textsc{Tomcat v6} was energy efficient on \dell and \serv, consuming 4.12\% and 4.3\% less energy, respectively.
We found that, for the same workload, systems running on \serv~consumed more than twice the energy they consumed on~\dell.

For both devices we can observe a trend of recommendations to replace well-known collections from the JCF (\texttt{Vector}, \texttt{ArrayList}, and \texttt{HashMap}) by alternative sources.
For the specific case of \textsc{Xalan}, among the 119 recommendations across the two desktop machines, just three suggested the use of implementations from the Java Collections Framework. 

\begin{figure}[tb]
	\centering
	\includegraphics[scale = 0.7, clip = true, trim= 3px 45px 0px 40px]{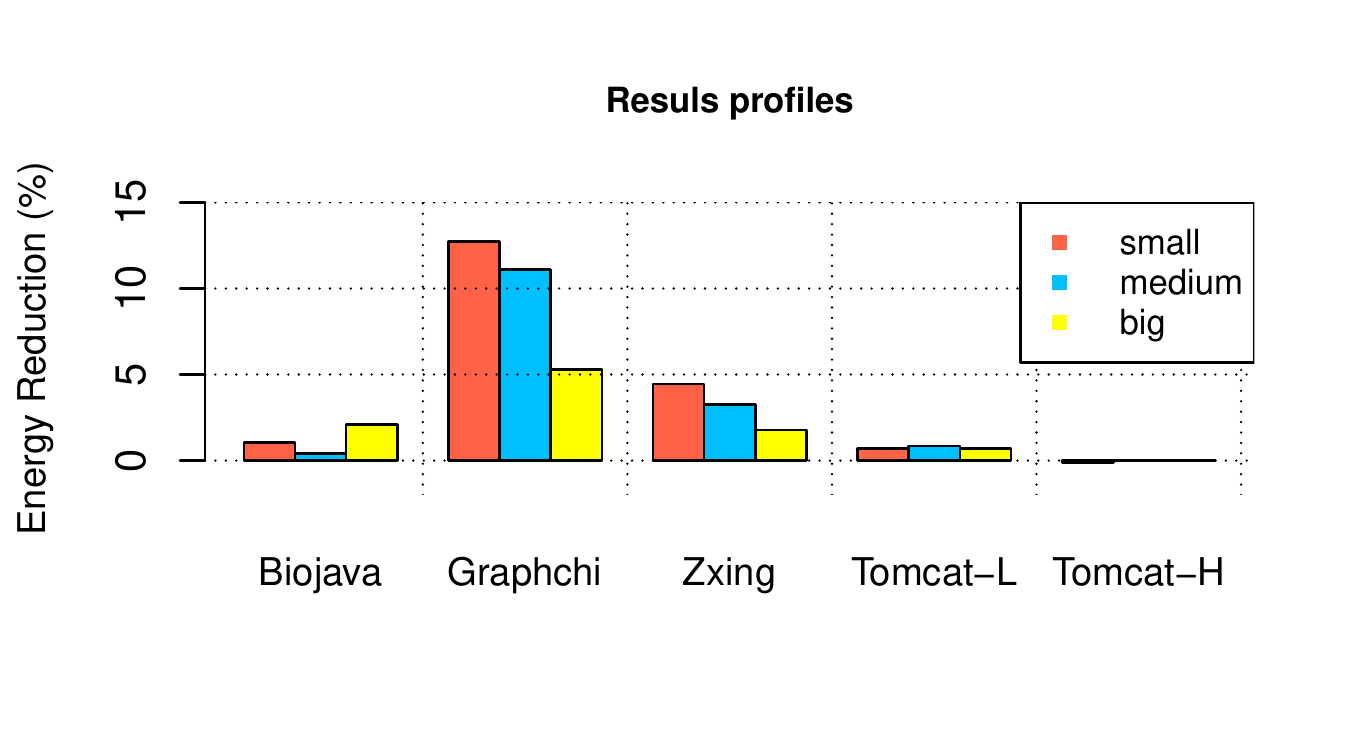}
	\caption{Percentage of energy reduction on the study about different profiles. Greater is better.}
	\label{img:resultsProfiles}
\end{figure}

On the mobile environment, \ct made 107 recommendations among the analyzed devices, with an expressive variation in their effectiveness. Modified versions of 
\textsc{PasswordGen} on \SEIGHT and \JSEVEN devices exhibited significant improvements over the original versions, consuming 4.7\% and 17.34\% less energy, but on \GTWO the modifications did not have significant impact.
The modified version of \textsc{Google Gson} exhibited an improvement of 5.03\% on \JSEVEN, however, the modifications yielded a small 0.95\% improvement on \SEIGHT. 
\textsc{Commons Math} had more inconsistent results. 
Although the modified version consumed 11.31\% less energy than the original version on \SEIGHT, that was not the case on \GTWO and \JSEVEN, where the modified versions consumed 1.2\% and 0.33\% \textbf{more} energy, respectively. 
Finally, \textsc{FastSearch} showed statistically significant results only on \SEIGHT, with the modified version having a very slight reduction of 0.09\% in energy consumption. 
\\[0.6cm]
\noindent
\textbf{Analyzing different profiles}.
Figure~\ref{img:resultsProfiles} summarizes the results of our modifications on the three different profiles.
Among the modified systems, \textsc{Graphchi} was the one which presented the best results, with a reduction of 12.73\% on \psmall, 11.09\% on \pmedium, and 5.30\% on \pbig.
While executing \textsc{Tomcat v9} using the \textsc{large} workload, \ct recommendations resulted in a reduction of energy consumed across all profiles.
On the other hand, while using the \textsc{huge} workload, \ct modifications did not provide a statistically significant result, with the exception of profile \psmall, where the modified version consumed more energy than the original version, making it less energy-efficient. 

Among the profiles, there was no overall winner. 
Each profile had at least one application with the best energy efficiency. 
\textsc{Tomcat-Large} consumes less energy using the recommendations made using the \psmall profile; \textsc{Graphchi}, and \textsc{Zxing} the \pmedium profile; and \textsc{Biojava} the \pbig profile.

There was an expressive variation in the number of implementation changes across the profiles, with \pmedium having the most with 456 changes, followed by \psmall with 352, and finally \pbig with 62, for a total of 870 recommendations that wielded reduced energy consumption. 

These changes were not evenly distributed, with some cases having a noticeable change in the number of recommendations made for an application based on the profile (e.g., \textsc{Biojava} had 13 times more changes on profile \pmedium than on profile \pbig). 
The changed collections were also different. As an example, the number of list implementations changed on \pbig are 5\% the number of list implementations changed on the other profiles.  
Also the changed collections were different. As an example, the number of list implementations changed on \pbig are 5\% the number of list implementations changed on the other profiles.  
Most of the time, these recommendations changed an implementation from JCF to an implementation from one of our alternative sources, i.e., Eclise Collections or Apache Commons Collections. More specifically, this was the case for 95\% of the recommendations when using the \psmall profile, 98\% for \pmedium, and 60\% for \pbig.

\subsection{Findings}\label{ss:collectionsDisc}

Analyzing more in-depth the results from both studies produced some interesting lessons.
\\[0.2cm]
\noindent
\textbf{Java Collections Framework is not the most energy-efficient.} The majority of the \ct recommendations were for collection implementations not in the JCF. In the desktop environment, only 5.8\% of the changes recommended by \ct used JCF implementations while in the mobile environment, \ct recommended JCF collection implementations in one third of the cases. 
Across the two different environments, 91.9\% of the recommendations originated from the Apache Common Collections and Eclipse Collections.
\\[0.2cm]
\noindent
\textbf{Collection popularity does not reflect energy-efficiency. } Looking at the most widely used collections in Java projects, i.e., \texttt{Hashtable}, \texttt{HashMap}, \texttt{HashSet}, \texttt{Vector}, and \texttt{ArrayList}, \ct changes to them comprise the best part of all changes made (94.2\% on the desktop environment and 94.39\% on the mobile environment).
In the specific case of \texttt{ArrayList}, the overall most popular Java collection by far~\cite{Oliveira:2019}, that happened for two reasons: First, two common operations (namely,  \texttt{insert(value)} and \texttt{iteration(random)}) usually have a worse performance in \texttt{ArrayList} when compared to other implementations. Second, in the cases where it was the best implementation, due to the its widespread use, it is already being employed and thus no benefits could be achieved.
\\[0.2cm]
\noindent
\textbf{The energy behavior varied heavily across devices}, even when executing the same application. 
Using \textsc{Xalan} as an example;
while being analyzed on \dell, \ct recommended ten \texttt{ArrayList} instances to be changed to \texttt{FastList} and one to \texttt{NodeCachingLinkedList}.
That was not the case on \serv, where \ct recommended only two instances of \texttt{ArrayList} and suggested the use of \texttt{TreeList}.
Nevertheless, there was an improvement in energy-efficiency in both machines. 
Another example is \textsc{Xstream}. 
All mobile modified versions, even while consuming less energy, did not differ statistically from their original version.
That was not the case on \dell, where the modified version consumed less energy and exhibited a statistically significant difference.
\\[0.2cm]
\noindent
\textbf{The profiles heavily influenced the energy savings.} 
Using the wrong profile can result in the energy consumption of the application rising instead of dropping, as in the case of the modified version of \textsc{Tomcat}, on the profile \psmall and using the workload with size \textsc{huge}.
That happened because \psmall was created to use small-sized collections,  
and thus it is optimized to that case while \textsc{huge} represents exactly the opposite. 
This illustrates that even though profile creation is an application-independent step of the proposed approach, knowledge about actual usage profiles can be leveraged to produce more useful energy profiles. 
A better use of the profiles can be seen in the recommendations applied to~\textsc{Tomcat} using the workload \textsc{large}, resulting in a positive impact on energy efficiency, with statistical significance in the three different profiles.
\\[0.2cm]
\noindent
\textbf{The best implementation is workload-dependent.} 
Among our recommendations on \asus, 95\% of list modifications on \psmall and \pmedium were changes from \texttt{ArrayList} to a different implementation. 
In \textsc{Biojava}, the system representing 60\% of all \texttt{ArrayList} modifications, two operations were the most intensively used: \texttt{insert(value)} and \texttt{iteration(iterator)}.  
This is reflected in the collection implementations that most often replace \texttt{ArrayList} in profiles \psmall and \pmedium, \texttt{NodeCachingLinkedList} and \texttt{FastList}, respectively, consuming less energy than \texttt{ArrayList} for these two operations. 
On the other hand, profile \pbig did not have a single implementation that had lower consumption for these operations, with \texttt{ArrayList} outperforming all the other implementations. 
Nevertheless, the changes made by \ct on \textsc{Biojava} resulted in an improvement in energy-efficiency across all profiles.
\\[0.2cm]
\noindent
\textbf{There is dominance between collections implementations}
Out of the 39 possible implementations available to \ct only 20 were recommended.
When trying to understand this behavior, we observed that some collection implementations consistently dominate~\cite{peterson:2009} others. 
A implementation $C_1$ dominates implementation $C_2$ when every operation in the former consumes less energy, on average, than the same operation in the latter. In this case, the dominated collection is never recommended by \ct.
Among all implementations, \texttt{ConcurrentHashMap} shows a particular behavior that is worth mentioning.
That implementation was changed 26 times on the desktop environment, 25 out of 26 cases for \texttt{ConcurrentHashMap(EC)}. 
Nevertheless, \texttt{ConcurrentHashMap} was also recommended 47 times, always replacing \texttt{Hashtable}.
This illustrates that if an implementation is not dominated by another, there will be cases where it may still perform better.
\\[0.2cm]
\noindent
\textbf{Energy profile creation is not trivial} 
During our experiments, we noticed that some factors could make it infeasible to create profiles at a larger scale. 
Due to the enormous variance in the execution times of operations, the original process~\cite{Oliveira:2019} of creating the energy profiles can take a long time, i.e., hours for desktop devices and days for mobile devices. 
To reduce that time we used two approaches:
First, we executed each operation three times, measured, and collected the energy consumption of those operations.
In the cases where a relation of dominance was found, the dominated collection was not included as an option for recommendation.
Second, we delimited a threshold based on how much time each operation could run.
This threshold was based on the fastest operation among all implementations in a specific group (e.g., \texttt{insert(start)} for thread-safe \texttt{List}s). Very expensive operations were discarded if they spent more time than our threshold. In our experiments, that threshold was set at two orders of magnitude, i.e., 100 times the average time of the fastest alternative.

\section{Energy Profiling in the Wild}\label{sec:greenhub}

Addressing energy efficiency within mobile devices is particularly relevant as these devices have become one of our most used gadgets, and  most often run powered by batteries. As a consequence, battery life is a high priority concern for users and one of the major factors influencing consumer satisfaction~\cite{FRichter2018,thorwartConsumer}. On the other hand, battery life is also important for app developers, as excessive battery consumption is one of the most common causes for bad app reviews in app stores~\cite{fu2013people,khalid2015mobile}. 

As we have already witnessed in the previous section regarding the choice of data structures, developer decisions can directly impact the energy consumption of a mobile application (or simply ``app''). In general, when considering other factors such as location services~\cite{lin2010energy}, programming languages~\cite{Oliveira:2017}, color schemes~\cite{Vasquez:2018:MOO,Wan:2017}, or code refactorings~\cite{couto2020energy}, the use of one available solution over another can have non-negligible effect on energy consumption. 

Keeping energy usage to a minimum is so important for app developers that IDEs for the most popular smartphone platforms include energy profilers. However, profiling for energy within mobile environments is a particularly difficult problem, and particularly within Android, the mobile platform with the largest market share, and by a big margin.\footnote{https://gs.statcounter.com/os-market-share/mobile/worldwide} Indeed, Android is a highly heterogeneous platform: in 2015 there were already more than 24 thousand Android device models available,\footnote{https://www.zdnet.com/article/android-fragmentation-there-are-now-24000-devices-from-1300-brands/} and a recent study found out that there are more than 2.5 million apps in the Google Play Store.\footnote{https://www.statista.com/statistics/276623/number-of-apps-available-in-leading-app-stores/} In addition, the Android operating system is currently in its 10th major release, with multiple minor releases throughout the years. These numbers combined with the different ways in which apps and devices are used produce a virtually infinite number of potential usage scenarios.

In this context, profiling for energy consumption has only limited applicability. Alternatively, one needs to obtain large-scale information about energy use in real usage scenarios to make informed, effective decisions about energy optimization.
In this section, we describe how we leverage crowdsourcing to collect information about energy in real-world usage scenarios. We introduce the GreenHub initiative, \textbf{\url{https://greenhubproject.org/}}, which aims to promote collaboration as a path to produce the best energy-saving solutions. The most visible outcome of the initiative is a large dataset, called Farmer, that reflects \textit{in the wild}, real-world, usage of Android devices~\cite{matalonga2019greenhub}. 

The entries in Farmer include multiple pieces of information such as active sensors, memory usage, battery voltage and temperature, running applications, model and manufacturer, and network details. This raw data was obtained by continuous crowdsourcing through a mobile application called BatteryHub. The collected data is strictly, and by construction, anonymous so as to ensure the privacy of all the app users. Indeed, it is impossible to associate any data with the user who originated it. 
The data collected by BatteryHub is then uploaded to a remote server, where it is made publicly available to be used by third parties in research on improving the energy efficiency of apps, infrastructure software, and devices.

In order to foster the involvement of the community,
Farmer is available for download in raw format and can be accessed by means of a backend Web app that provides an overview of the data and makes it available through a REST API. The dataset can also be queried by means of LumberJack, a command-line tool for interacting with the REST API. 

Within the GreenHub initiative, we were so far able to collect a dataset which is \textit{sizable}. Thus far it comprises 48+ million unique samples. The dataset is also \textit{diverse}. It includes data stemming from 2.5k+ different brands, 15k+ smartphone models, from over 73 Android versions, across 211 countries.

In the remainder of this section, we describe in  detail the alternatives that we have implemented to allow the community to access the data within our dataset. The main motivation for the section is to foster the engagement of the community in exploring the data we are providing, in this way contributing to increase the knowledge on how energy saving strategies can be realized within Android devices.
We therefore expect that this section is particularly interesting for researchers and/or practitioners focusing on mobile development, and that have a data mining mindset.

\subsection{A collaborative approach to Android energy consumption optimization}

The success of GreenHub is dependent on its data, and to keep such data coming in, we plan to give back to the community in concrete and valuable ways. In this section, we focus on the ways that the community can access the data we are collecting.

\normalsize
\begin{figure}[tb]
\centering
\includegraphics[width=0.5\columnwidth]{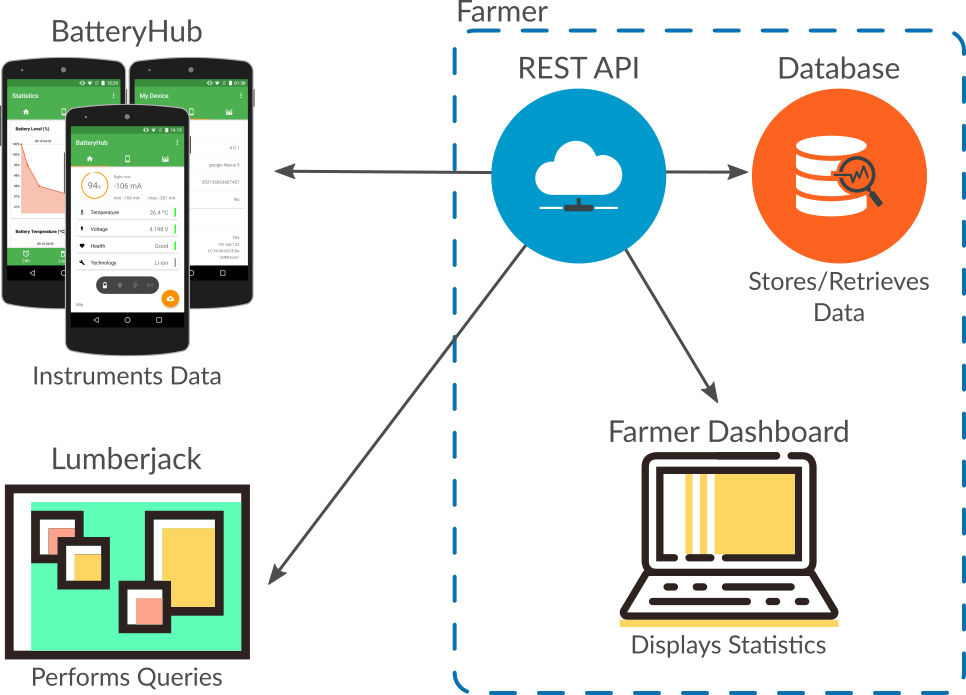}
\caption{GreenHub platform architecture. }
\label{fig:greenhubInfra}	
\end{figure}

The initiative relies on an open-source technological platform\footnote{\url{https://github.com/greenhub-project}}, whose architecture overview is shown in Figure~\ref{fig:greenhubInfra}. This platform includes our data collection Android app called BatteryHub, a command-line application interface called Lumberjack, and the Farmer REST API for prototyping queries, dashboard interface, and database for storing data. These components are further defined in the following sub-sections.
\\[0.6cm]
\noindent
\textbf{Data Collection} is provided by BatteryHub, an Android app whose development was inspired by Carat~\cite{oliner2013carat}. Initially, we forked Carat's open-source code to take advantage of the data collection and storage mechanisms. On top of that, we  updated its data model to consider more details on modern devices, such as NFC and Flashlight usage, for example. In the same spirit of Carat, BatteryHub is entirely open-source. In contrast with Carat, however, all our collected data is permanently and publicly available, so as to strongly encourage and help others in collaborating, inspecting and/or reusing any artifact that we have developed or collected.

BatteryHub is available at Google's Play Store\footnote{\url{https://play.google.com/store/apps/details?id=com.hmatalonga.greenhub}}, and tracks the broadcast of system events, such as changes to the battery's state and, when such an event occurs, obtains a sample of the device's current state. 
Batteryhub either uses the official Android SDK or custom implementations for universal device compatibility support, and periodically communicates with the server application (over HTTP) to upload, and afterwards remove, the locally stored samples.
Each sample characterizes a wide range of aspects that may affect battery usage, such as sensor usage, temperature, and the list of running applications. 

It is important to mention that the data collected from each user is made anonymous by design.
Each installation of BatteryHub~is associated with a random unique identifier and no personal information, such as phone number, location, or IMEI, is collected. This means that it is (strictly) not possible to identify any BatteryHub user, nor is it possible to associate any data with the user from whom it originates. 

In regards to sample collection frequency, a new data measurement is collected (to be sent to our server) when the battery's state changes. In most cases, this translates to a sample being sent at each 1\% battery change (which accounts to 95\% of the time according to our data).
The app allows for configurable alerts, e.g. when the battery reaches a certain temperature. Our overarching goal is to use BatteryHub to give suggestions to users, based on their usage profiles, on how to reduce the energy consumption of their device.

Besides BatteryHub, our infrastructure includes four additional components, as depicted in Figure~\ref{fig:greenhubInfra}. We envision they can be used in different stages of mining our dataset, which are described in the remainder of this section.
Finally, our infrastructure also includes a web dashboard interface\footnote{\url{https://farmer.greenhubproject.org/}}
that provides access to up-to-date statistics about the collected samples.
\\[0.6cm]
\noindent
\textbf{Fast prototyping of queries}
can be made by using Farmer's REST API, which was designed as a means to quickly interface with and explore the dataset. As every request made to the API must be authenticated, users must first obtain an API key in order to access the data in this fashion\footnote{\url{https://docs.greenhubproject.org/api/getting-started.html}}.  
The API provides real-time, selective access to the dataset and one may query, e.g., all samples for a given brand or OS version. Since the API is designed according to the REST methodology, this allows us to incrementally add new data models to be reflected within the API itself as the data protocol evolves over time.
After an API key has been successfully generated, one may request his/her own user profile from the API:

{\small
\begin{verbatim}
farmer.greenhubproject.org/api/v1/me?api_token=yourTokenHere
\end{verbatim}
}

\noindent 
Every successful API response is a JSON formatted document, and in this case the server will reply with the user details, as shown next.

{\scriptsize
\begin{verbatim}
{   "data": {
        "id": XX,
        "name": "Your Name",
        "email": "your@email.com",
        "email_verified_at": "YYYY-MM-DD HH:MM:SS",
        "created_at": "Mon. DD, YYYY",
        "updated_at": "YYYY-MM-DD HH:MM:SS",
        "roles": [ ... ]   }   }
\end{verbatim}
}

\noindent
It is now possible to use the API, for example, to list devices:

{\small
\begin{verbatim}
farmer.greenhubproject.org/api/v1/devices?api_token=yourToken
\end{verbatim}
}

\noindent
This request can take additional parameters for \textit{page} and devices \textit{per page}. A full description of all the available parameters for each request can be found in the API Reference\footnote{\url{https://docs.greenhubproject.org/api-reference/}}. 

The expected response from the request above is as follows: 

{\scriptsize
\begin{verbatim}
{   "data": [
	    {   "brand": "asus",
            "created_at": "2017-10-28 02:51:09",
            "id": 2518,
            "is_root": false,
            "kernel_version": "3.1835+",
            "manufacturer": "asus",
            "model": "ASUS_X008D",
            "product": "WW_Phone",
            "os_version": "7.0",
            "updated_at": "2017-10-28 02:51:09"
        },  ... ],
    "links": { ... },
    "meta":  { ... }   }
\end{verbatim}
}

\noindent
To get more detailed information, e.g., about a particular device whose identifier is 123, it is possible to request its samples:

{\scriptsize
\begin{verbatim}
farmer.greenhubproject.org/api/v1/devices/123/samples?api_token=yourToken
\end{verbatim}
}

A complementary approach to interface with the API is to use its command-line application interface, Lumberjack.
Using this tool, users can perform flexible, on-demand queries to the data repository, to support quick prototyping of data queries applying different filters and parameters. Furthermore, users can quickly fetch subsets of the data without the need to download a snapshot of the entire dataset. The following is an example of a Lumberjack query to obtain the list of Google brand devices:

{\scriptsize
\begin{verbatim}
  $ greenhub lumberjack devices brand:google -o googleDevices.json
\end{verbatim}
}

\noindent
The following example queries the dataset for samples whose model is nexus and that were uploaded before May 31st 2018: 

{\scriptsize
\begin{verbatim}
  $ greenhub lumberjack samples model:nexus -R ..2018-05-31     
\end{verbatim}
}
\noindent
\textbf{Extensive Mining} can be conducted from the samples we collected, and which are accessible through Farmer. The dataset is available as a zip archive file, in CSV format\footnote{\url{https://farmer.greenhubproject.org/storage/dataset.7z}}, and also in Parquet\footnote{\url{http://parquet.apache.org}} binary format\footnote{\url{https://farmer.greenhubproject.org/storage/dataset.parquet.7z}}, which can be analyzed more efficiently than a plain text dump. 
The dataset is also available as a MariaDB relational database. The samples sent by BatteryHub are queued to be processed by a PHP server application built using the Laravel framework\footnote{\url{https://laravel.com/}}. Each sample is received as a JSON formatted string that is deconstructed and correctly mapped within the database. 

\begin{figure}[htb!]
\centering
\includegraphics[width=0.6\columnwidth]{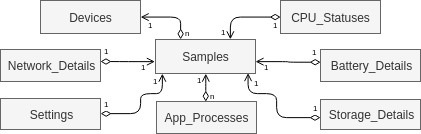}
\caption{Entity Relationship diagram of the dataset}
\label{fig:datamodel}	
\end{figure}

The (simplified) data model that we employ is shown in Figure~\ref{fig:datamodel}, where each box represents a table (or a CSV file) in the dataset. \textsf{Samples} is the most important of them, including multiple features of varied nature, e.g., the unique sample id, the timestamp for each sample, the state of the battery (charging or discharging), the level of charge of the battery, whether the screen was on or not, and the free memory on the device.

\textsf{App\_Processes} is the largest among the tables of the dataset, containing information about each running process in the device at the time the sample was collected, e.g., whether it was a service or an app running on the foreground, its name, and version. \textsf{Battery\_Details} provides battery-related information such as whether the device was plugged to a charger or not and the temperature of the battery. \textsf{Cpu\_Statuses} indicates the percentage of the CPU under use, the accumulated up time, and sleep time. 
\textsf{Devices} provides device-specific information, such as the model and manufacturer of the device and the version of the operating system running on it. \textsf{Network\_Details} groups network-related information, e.g., network operator and type,  whether the device is connected to a wifi network, and the strength of the wifi signal. The \textsf{Settings} table records multiple yes/no settings for services such as bluetooth, location, power saver mode, and nfc, among others. Finally, \textsf{Storage\_Details} provides multiple features related to the secondary storage of the device.

\section{Conclusion}

Although developers have recently become more aware of the importance of creating energy-efficient software systems, they are still missing important knowledge and tools to help them achieve their goals. In summary, with this chapter we hope to show the following relevant aspect of green development:
\\[0.2cm]
\noindent
\textbf{Small changes can make a big difference in terms of energy consumption}, especially in mobile devices. 
Throughout this chapter we discussed the importance of diversity and the energy consumption of different pieces of software, such as I/O APIs, concurrent mechanisms, and Java collections.
For every one of these constructs, performing modifications to employ diversely-designed, more energy-efficient versions resulted in a reduction in energy consumption, with some cases only requiring changing a single line of code.
Focusing on these energy variation hotspots can greatly reduce the complexity of improving the energy-efficiency of an application, making it viable for non-specialists to enhance their systems.
Knowing this, developers and researchers can focus on easily-exchangeable, power-hungry aspects of the software to reduce the energy consumption of an application with minimum effort.
\\[0.2cm]
\noindent
\textbf{Device variability is a real problem when experimenting on energy consumption}.
Similarly to performance, energy efficiency can be impacted by a number of factors, at different abstraction levels, including factors that are not obvious for non-specialists.
Factors such as a device's battery's age or the room temperature can have a significant influence on the energy consumption.
Even worse, when dealing with different devices, every decision made by the manufacturer can change the energy-efficiency of the device and turn a seemingly sound experiment into conclusions that only work in specific cases.
Knowing this, developers and researchers can try to mitigate this factor by executing their experiments on bigger pool of devices, greatly reducing this bias factor on their results. 
In case only a single device is available, its characteristics should be fully presented so other researchers/developers can have a clear understanding of the results and in which devices they would be relevant.
\\[0.2cm]
\noindent
\textbf{Crowdsourcing can be used to see the big picture of energy consumption}.
Although very important for several factors, such as studying particular cases and learning about energy consumption behavior, controlled experiments on mobile devices hardly extrapolate to the whole environment.
Even when dealing with just the Android OS, there is so much variability (e.g., OS versions, manufacturers, device versions) that no controlled experiment will be truly general enough to cover a significant number of devices and thus have widespread  applicably.
Crowdsourcing can be a way to mitigate that by using the data provided by the users, leveraging thousands of different devices to achieve a truly panoramic view of the ecosystem as a whole.
The biggest problem with the crowdsourcing solution is that it depends on end-users for the energy samples. 
Convincing them to provide their data and, even more importantly, getting enough of it to be statistically relevant, can be arduous. 
We propose that developers and researchers that want to investigate the energy behavior of the mobile devices in the Android environment to use the GreenHub initiative, an already well-established database with millions of data-points.   

\bibliographystyle{plain}
\bibliography{biblio}

\section*{Author's Biographies}

{\small 
\authorbio{\textbf{Wellington Oliveira}} is a PhD student at the Federal University of Pernambuco. He is member of Software Productivity Group and is currently researching in the field of green computing and experimental software engineering. More specifically, he does his research on how to use static analysis to automatically optimize the energy efficiency through modifications in the application source code. 

\authorbio{\textbf{Hugo Matalonga}} is a MSc student at the Minho University. His research interest is in the field of green computing and explainable AI. He co-founded the GreenHub project and has been the Lead Developer of the project since the very beginning. Currently he is working on studying energy consumption in Android devices which factors and how different environments impact the energy usage patterns applying statistical models.

\authorbio{\textbf{Gustavo Pinto}} is an Assistant Professor at the Federal University of Pará. He holds a Ph.D. in Computer Science from the Federal University of Pernambuco in 2015. He works on the intersection between people and code, approaching topics such as refactoring, human aspects of software development, contributions and contributors to open source software, and, obviously, energy efficiency.

\authorbio{\textbf{Fernando Castor}} is an Associate Professor at the Informatics Center of the Federal University of Pernambuco, Brazil. His broad research goal is to help developers build more efficient software systems more efficiently. More specifically, he conducts research in the areas of Software Maintenance, Software Energy Efficiency, and Error Handling.

\authorbio{\textbf{Jo{\~a}o Paulo Fernandes}} is an Assistant Professor at the
Informatics Engineering Department of the University of Coimbra. He holds a
Ph.D. in Computer Science from the University of Minho, Portugal. He has done
extensive work towards improving the energy efficiency of the software
components of IT systems, having founded GreenHub, a
crowdsourcing initiative that aims to optimize and reduce the power
consumption associated with the use of
mobile devices, namely within Android. He has also co-founded the GSL - Green
Software Laboratory and the GreenHaskell - Towards Energy-Efficient Programming
in Haskell research projects. Both projects aim to produce knowledge and develop
tools that can assist developers in their need and ambition to develop software
products that are more environmentally sustainable. Jo{\~a}o Paulo Fernandes has
described his research achievements in 60+ publications in top tier venues such
as IEEE TSE, JSS, ICSME, MSR, SANER or SLE.

}

\end{document}